\documentclass[ahp]{essaibirk}

\usepackage{amsmath,amstext,amssymb}
\usepackage{graphicx}

\begin{document}

\newcommand{\Wr}{\overrightarrow{\omega}}
\newcommand{\Wl}{\overleftarrow{\omega}}
\newcommand{\Ur}{\overrightarrow{\rho}}
\newcommand{\Ul}{\overleftarrow{\rho}}
\newcommand{\be}{\begin{equation}}
\newcommand{\ee}{\end{equation}}

\title[Fluctuation Relations for Molecular Motors]{Fluctuation Relations for Molecular Motors}

\author[D. Lacoste]{David Lacoste}

\address{Laboratoire de Physico-Chimie Th\'eorique, \\ UMR 7083 CNRS\\ 10, rue Vauquelin\\ 75231 Paris Cedex 05, France}
\email{david.lacoste@gmail.com}

\author[K. Mallick]{Kirone Mallick}

\address{Institut de Physique Th\'eorique\\ CEA Saclay\\ 91191 Gif sur Yvette, France}
\email{Kirone.Mallick@cea.fr}

\begin{abstract}

This review is focused on the application of specific fluctuation
relations, such as the Gallavotti-Cohen relation, to ratchet
models of a molecular motor. A special emphasis is placed on
two-states  models such as the flashing ratchet model. We derive
the Gallavotti-Cohen fluctuation relation for these models and we
discuss some of its implications.
\end{abstract}
\maketitle


\section{Introduction}

    The macroscopic observables of a  system at mechanical and  thermal equilibrium do not vary with time
 and can be characterized by a finite number of state variables. Thermodynamics imposes {\it a priori} constraints
 on the average values of these  state variables that are  satisfied regardless of the specific nature of the system.
 Because of this  property of time-invariance, equilibrium is often  imagined as being
 associated with   stillness  and frozen dynamics. This, of course, is not true:
 a system, even at   thermodynamic equilibrium,  is constantly  evolving   from one
 micro-configuration to another. This endless motion at the microscopic level can be probed macroscopically
 by measuring the fluctuations of some physical observables,  the most famous example being Brownian Motion.
 Equilibrium fluctuations  are perfectly  well explained
 by the classical laws of  statistical mechanics.

Brownian motion, its nature, its origins,   have
 been a puzzle to physicists during the XIXth century \cite{duplantier:2006}.
One  paradoxical  issue  was the following: can one rectify  this random fluctuating motion
and use it to perform  some work ? If  such a  rectifying device
could be constructed then  work would be extracted  from a single heat reservoir,
contradicting   the second law of thermodynamics.
 The most famous  example of a mechanical system that may play  the  role of such  a Maxwell's demon is
 the ratchet and pawl system, presented by Feynman in Chapter 46 of the
 first volume of  his Physics Lectures \cite{feynman}.
 A related paradox was proposed in 1950 by Brillouin \cite{brillouin:1951}:
 consider an electrical circuit composed of a diode and a resistor at temperature $T$. The current in the
 circuit has a zero average value, but because of thermal noise, it exhibits non-vanishing
 fluctuations (known as Johnson-Nyquist noise).
 Can the diode be used to rectify the current, allowing us to use it  to  perform  work?
 The solutions of these paradoxes is  now well-known: thermal fluctuations are a
 universal phenomenon  and all systems at a given temperature are subject to it. The rectifying device,
 whatever it is,
 is also subject to Brownian motion and  undergoes some  unavoidable  fluctuations. If the
 signal to be rectified  is produced at temperature $T$ and the  rectifier is at the same temperature then the
 thermal  fluctuations of the rectifier render  it totally ineffective   and the second law is saved.
 If the rectifier is at a lower temperature then  this apparatus can indeed generate  work:  but now
 there are  two heat sources at  different temperatures, in accordance  with  the second law.

 In recent years, a renewed interest has arisen in ratchet
  models in the context of non-equilibrium
 statistical physics. Many ratchet models exist (for a review see \cite{reiman}).
In one kind of ratchet called brownian motors, an association of 
two particles, one asymmetric and another one symmetric, can
rectify thermal fluctuations provided that the two particles are in contact with
heat baths at different temperatures \cite{broek:2005}. 
In another family of ratchet models, the 
need of two heat baths is removed by coupling the ratchet
 to some external `agent'
  (e.g.,  a chemical reaction) that continuously
 drives the system out of equilibrium. In this case too,
under certain conditions, work can be extracted \cite{magnasco1,ajdari:1993,astumian:1997}.
 Again, there is no contradiction with
 thermodynamics here: the system is far from equilibrium
 and the ratchet plays simply the role of a transducer
  between the energy put in by the agent (e.g. chemical
 energy) and the mechanical work extracted.
The analysis of the energetics of such  devices far from
 equilibrium requires concepts that go beyond the classical
laws of thermodynamics and this remains
 a very challenging and important open issue \cite{ken:1997,parrondo:2002}.

 Biophysics provides numerous examples
 of systems far from equilibrium. For example,
  a significant part of the eukaryotic cellular traffic relies
 on 'motor' proteins that move  along filaments
 similar in function to railway tracks or freeways (kinesins and dyneins
 move along tubulin filaments; myosins move along actin filaments) \cite{howard}. The
 filaments are periodic (of period $\sim 10$nm)  and have a fairly
 rigid structure; they are also polar: a given motor always moves in the
 same direction. These molecular   motors appear in a variety of biological contexts:
  muscular contraction, cell division, cellular traffic, material
 transport along the axons of nerve cells...
 A biological cell forms a crowded environment in which
 molecular motors work together and with other proteins. In these
conditions, collective effects arise due to interactions between motors \cite{gautam:1999}.
In many cases, these interactions can be modeled as excluded volume interactions,
and for this reason, the behavior of an ensemble of motors in a low dimension space can
be described by dynamical models similar to the ones developed for traffic problems
\cite{chowdury:2008,mueller:2008}. In the following, we focus on
single molecular motor properties, in order to clarify in this simpler situation, the dynamics and the energetics of this system far from equilibrium.

Recently, a general organizing principle for non-equilibrium
systems has\break emerged which is known under the name of fluctuation
relations
\cite{GC:1995,Jarzynski:1997,kurchan,evans,crooks,Jarzynski:2008,ritort-2008}.
These relations, hold for non-equilibrium steady states but
arbitrarily far from equilibrium
 \cite{lebowitz,derrida-2007,qian-2001,qian-2005}, they can be
seen as macroscopic consequences of the invariance under time
reversal of the dynamics at the microscopic scale
\cite{chetrite-CMP:2008}. It is interesting to apply these
concepts to small systems which can either be mechanically driven
as biopolymers \cite{ritort:2005} or chemically driven as enzymes
\cite{schmiedl-2007}.

Molecular motors are enzymes which operate
stochastically at the level of a few molecules, and for this
reason they typically undergo large thermal fluctuations.
Generically,
single molecular motors have been described theoretically either by
continuous ratchet models (see e.g. review by J\"ulicher et al.
\cite{armand1})  or by models based on master equations on a discrete space
\cite{kolomeisky-revue,lipowsky-2000}.
It is possible to give a thermodynamic interpretation of the elementary chemical
reactions, which occur in a discrete and stochastic way in a
molecular motor \cite{seifert}. Such a thermodynamic
interpretation of chemical transitions has similarities with the
thermodynamic interpretation of the Langevin equation at the
single trajectory level \cite{ken:1998}. At a macroscopic scale,
constraints arise on the operation of these molecular motors, as a
result of single reaction events occurring stochastically at the
microscopic scale. These constraints take the form of a
fluctuation relation \cite{gaspard2,seifert,lipowsky:2008,
prl-FT,pre-FT,maes}.

 The aim of this review is to explain how recent
 theoretical  results in non-equilibrium statistical
 mechanics, namely these fluctuation relations,
 provide a way to understand
 the non-equilibrium energetics of
 molecular motors. Note that the same framework apply to both biological
 molecular motors or artificially made nanomachines.

In the first part of this review, we present the two theoretical
approaches of molecular motors mentioned above, namely the
flashing ratchet model and the  approach based on
master equations on a discrete space.
In the second part of the review, we  derive  the fluctuation
relations  for these specific models  and
 discuss some consequences of these relations
 for   molecular motors.

\section{Stochastic models of molecular motors}

 Molecular motors are enzymes capable of converting chemical energy
derived from the hydrolysis of adenosine triphosphate (ATP) into
mechanical work. There is a large diversity of molecular motors,
and correspondingly a large number of processes accomplished by
 these  motors within a cell. There are linear motors such as
kinesins, dyneins, myosins or the RNA polymerases, and rotating motors
such as the $F_1$-ATPase motor or the bacterial flagellar motor.
These motors drive not only
intracellular movements, they are also key players in the motility
of the cell itself. Although, traditionally, these machines were
subjects of investigation in biology and biochemistry, increasing
use of the concepts and techniques of physics in recent years have
contributed to a  quantitative understanding of the fundamental
principles of operation of these motors. The possibility of
exploiting these principles for the design of artificial
nanomachines has opened up a new field in nanotechnology.

  \subsection{The basic  principle}

\begin{figure}[ht]
  \includegraphics[height=7.5cm]{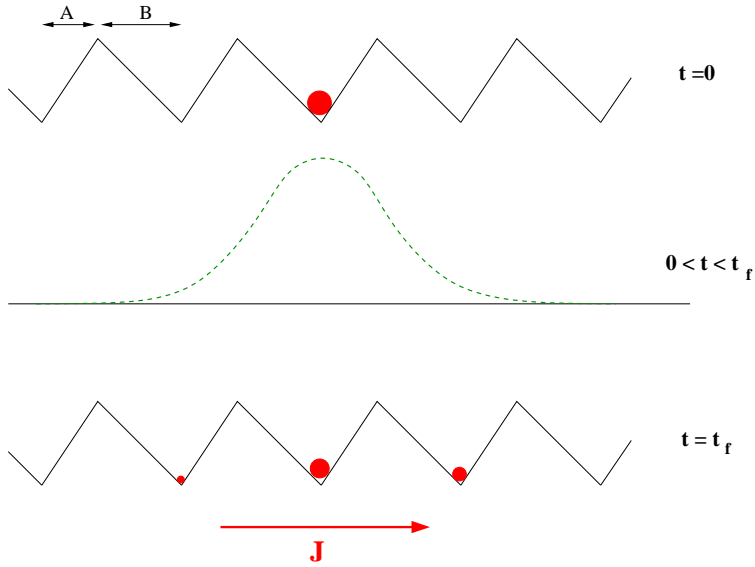}
\caption{The principle of a Brownian ratchet:
 by inscribing  and erasing periodically
 an  asymmetric potential, a directed motion of the particle is induced.
 In this example, the potential is a saw-tooth function of period $a= A +B$.
 Since  $B > A$,  the potential is   asymmetric.
 The relative sizes
 of the  probabilities of ending in one of the wells
  are represented by the sizes of the disks in the
 lowest picture.
  The  right and left  probabilities being
  different, this  leads on average to a net total current $J$.}
  \label{fig:MotorMol2}
\end{figure}

On the theoretical side, molecular motors have been described by ratchet
devices, which are systems able to extract useful work out of unbiased random fluctuations \cite{hanggi}. A generic
model of such a ratchet device is shown in Figure~\ref{fig:MotorMol2}
 where the motor is represented
 by a small particle that  moves in a one-dimensional space.
  At the initial time $t=0$,  the motor is trapped in
 one of the wells of a periodic  asymmetric
  potential of period $a$.
  Between time 0 and $t_{f}$, the asymmetric  potential is erased
 and the   particle diffuses freely and isotropically at temperature $T$.
  At the switching  time $t_{f}$, the  asymmetric potential is
 re-impressed, the motor slides down to the nearest  potential valley
  and, because of damping,  is  trapped in
  one  of the wells.
  The motor has maximal chance
 to end up  in the same well where it was at time $t=0$.
 However,  it has a small
 probability to be trapped in the  well located to the right and,
  because of the  asymmetry of the potential,  an even
  smaller  probability to end up  in the left  well.
 Indeed, in order to be trapped in the right well after time $t_{f}$,
 the particle must have diffused between $t=0$  and $t= t_{f}$  over a distance larger than  $A$ towards
  the right. However,  to end up  in the left  well it has to diffuse (towards the  left)
 a distance  larger than  $B$, which is much less probable because $B>A$.
  In other words,  because the potential is asymmetric,
  the motor has more chances to slide down towards the right: this
 leads on average to a net total current. The particle has used thermal noise
 to overcome the potential barrier and thanks to the  asymmetry of the potential it has moved
 in a well-defined direction.

\begin{figure}[ht]
  \includegraphics[height=3.5cm]{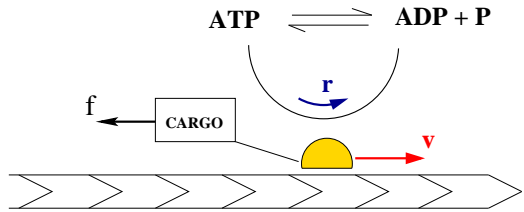}
 \caption{Schematic representation of a molecular motor: by hydrolyzing  ATP,
 the motor  proceeds along the polar filament  and carries a 'cargo' molecule,
 which typically exerts a force $f$ on the motor.}
  \label{fig:MotorMol1}
\end{figure}

In order to move the motor consumes $r$ ATP
 fuel molecules per unit time, which are hydrolyzed to ADP + P  (see Fig.~\ref{fig:MotorMol1}):
 $$  ATP \rightleftharpoons  ADP + P  \, .$$
  It is the chemical energy released by  ATP-hydrolysis that allows the motor to detach itself
 from the filament it was  bound to. This detachment process corresponds   in the  basic mechanism  to   erasing
 the  potential  whereas re-attachment of the   motor  at a the switching
 time $t_{f}$ corresponds to  re-impressing
 the  potential. Hence the motor undergoes chemistry-driven changes between strongly and  weakly
 bound states (attachments and detachments). It is the coupling between chemistry and the interaction
 with the filament that allows the motor to overcome energy barriers. Besides, because of the polarity
 of the filament, the interaction  potential is asymmetric, allowing directed motion to set in.

 In general, the motor is subject to an external
  force $f_{{\rm ext}}$ which tilts the potential.
  Besides, when ATP is in excess, the chemical potential difference
  of the reaction of ATP hydrolysis,
 $\Delta \widetilde{\mu}  =\mu_{{\rm  ATP}} - \mu_{{\rm  ADP}} -\mu_{{\rm P}}$
  becomes {\it positive}.
More precisely, we denote $\Delta \widetilde{\mu} \equiv k_B T \Delta \mu$, where
$\Delta \mu$ is the normalized chemical potential and
\begin{equation}\label{def of delta mu}
\Delta \tilde{\mu} = k_B T  \ln \left( \frac{[ATP] \, [ADP]_{eq}
\, [P]_{eq}} {[ATP]_{eq}\, [ADP] \, [P]} \right),
\end{equation}
where $[..]$ denotes concentration under experimental conditions
and $[..]_{eq}$ denotes equilibrium concentrations.

 A basic problem  is then  to  determine  the  {velocity of the motor}
  $v(f_{{\rm ext}}, \Delta\mu)$  (mechanical
 current)  and the {ATP consumption rate}
  $r(f_{{\rm ext}}, \Delta\mu)$  (chemical current)
  as functions of the external
   mechanical and chemical loads.

To summarize, molecular motors move by using the ratchet effect, providing  an example
 of a  {\it rectification process} of Brownian motion.  The two basic requirements for obtaining
 directed motion are:

 (i)  an external energy source,
  provided by the chemical reaction of  ATP hydrolysis. During this
 process  ATP is consumed and ADP is produced. This reaction  therefore
 breaks time-reversal invariance (more technically, it breaks the detailed balance
 condition which holds at equilibrium, and introduces a bias in the dynamics of the
 motor).

(ii) The polarity of the filament which breaks spatial left-right symmetry
 and allows the motor to move in a well-defined direction.

\subsection{The flashing ratchet model}

 In the simplified discussion above, we did not specify the characteristics
 of the switching time $t_f$. Different models are possible: one can consider
 a deterministically forced ratchet in which the binding potential
 of the motor switches periodically (and even smoothly)  from strong-binding
 to weak-binding. Another possibility, that
  we shall discuss here  in detail is the flashing ratchet model \cite{armand1,reiman}
 in which the switching of potentials  is sudden  and  occurs at  random times generated
 by a Poisson Process.
  Then,  we shall  show how to construct a  discrete version  of
  flashing ratchet model \cite{kolomeisky-revue,lipowsky-2000}.

In the flashing ratchet
model \cite{armand1,parmeggiani,elston}, the state of the motor is described
by a continuous position variable
 $x$ and by discrete internal states $i=1,2$ corresponding to
 different chemical states of the motor. For instance,  one could
associate one state with a configuration where one motor head is bound
 to the filament (the high energy state) and the other state
to a configuration where both heads are bound (the low energy state).
The motor evolves in two time-independent periodic potentials $U_i(x)$, with
$i=1,2$. Note that in the basic mechanism discussed
 in the previous section,  $U_1(x)$ is a saw-tooth potential and
 $U_2(x)$ is  taken to be zero. But  one can consider  the general  case where
  both  $U_1$ and  $U_2$
 are non-zero asymmetric potentials  of arbitrary shape  with  a common period $a$.
 In figure \ref{fig:ratchet}, we represent the  often studied
 situation in which  $U_1$ and  $U_2$ are identical  saw-tooth potentials
 but slightly shifted  with respect to each other along the $x$ axis.

 The dynamics of the motor can be
 represented by a Langevin equation
 \begin{eqnarray}
 \dot x =   -\gamma F  -\gamma \sum_{i=1,2}  U_i(x) \delta_{\zeta(t),i} \, +    \sqrt{D_0} \, \xi(t)
 \end{eqnarray}
 where $\xi(t)$ is a normalized white-noise and $\zeta(t)$
 a dichotomous noise that can exist in two states 1 and 2.
 The switching-rates of  $\zeta(t)$ are position dependent and  are given by
  $\omega_1(x)$ (transition from 1 to 2)  and $\omega_2(x)$ (transition from 2 to 1).
 The friction coefficient $\gamma$ satisfies the Einstein relation
  $ D_0 = k_BT/ \gamma $ and $F$ represents an external force
 acting on the motor. The function $ \delta_{\zeta(t),i}$ is a Kronecker delta.

The probability density for the motor to be at position
 $x$ at time $t$ and in state $i$ is denoted by  $P_i(x,t)$,
  which  obeys the following equations
 \begin{eqnarray}
\label{eqs:moteur1} \frac{\partial P_1}{\partial t }  +
\frac{\partial J_1}{\partial x}
 &=& - \omega_1(x)  P_1 +   \omega_2(x)  P_2     \\
\label{eqs:moteur2}
\frac{\partial P_2}{\partial t }  +
\frac{\partial J_2}{\partial x}
 &=&  \omega_1(x)  P_1  -   \omega_2(x)  P_2  \, ,
 \end{eqnarray}
where $\omega_1(x)$ and $\omega_2(x)$ are space dependent
transition rates, and the local currents $J_i$ are defined by:
\begin{equation}
      J_i = -D_0 \left(  \frac{\partial P_i}{\partial x}
         +
   \frac{1}{k_B T} \left( \frac{\partial U_i}{\partial x} - F \right) P_i \right),
\end{equation}
with $D_0$ the diffusion coefficient of the motor
and $F$ a non-conservative force acting on the motor.

The transition rates can be modeled using standard kinetics for
chemical reactions  applied to each chemical pathway between the
two states of the motor \cite{parmeggiani}:
 \begin{eqnarray}
        \omega_1(x) &=& [ \omega(x) + \psi(x) e^{\Delta\mu} ] e^{(U_1(x)-fx)/k_B T}, \nonumber \\
        \omega_2(x) &=& [ \omega(x) + \psi(x) ] e^{(U_2(x)-fx)/k_B T},
        \label{rates omega}
\end{eqnarray}
where $f=F a / k_B T$ is
the normalized force acting on the motor.
It is assumed that the rates can be decomposed into a contribution proportional to $\omega(x)$, which is associated with thermal transitions, and a contribution proportional to $\psi(x)$ corresponding to transitions induced by ATP hydrolysis. Note that the functions $\omega(x)$ and $\psi(x)$
have to be periodic functions but they are otherwise unspecified. The form of the rates in
the absence of hydrolysis ({\it i.e.} when $\psi(x)=0$) is chosen to satisfy the detailed balance condition
\begin{equation}
    \frac{ \omega_2(x)} {\omega_1(x)} =
  \exp\left( \frac{U_2(x) - U_1(x)}{k_B T}\right) \, .
  \label{usual DB}
\end{equation}
The form of the rates associated with the transitions induced by
ATP hydrolysis ({\it i.e.} when $\omega(x)=0$) is chosen to
satisfy a generalized detailed balance condition, which is
generalized in the sense that it accounts for the exchange of
chemical energy \cite{lipowsky:2008,schmiedl-2007}. In this case,
this leads to the condition
\begin{equation}
    \frac{ \omega_2(x)} {\omega_1(x)} =
  \exp\left( \frac{U_2(x) - U_1(x)}{k_B T}- \Delta \mu \right) \, .
  \label{generalized DB}
\end{equation}
 Note that the way the force enters the rates is unambiguous in continuous models as compared
  to discrete models, in which the force dependant rates must contain unknown load distribution factors
   \cite{pre-FT,kolomeisky-revue}.
One could easily extend the model to introduce more chemical pathways \cite{parmeggiani}
or more internal states; such extensions are possible but they have not been considered here since
they are not essential for the present argument.

When $F=0$ and $\Delta \mu=0$, the system is in equilibrium since the detailed balance condition
(\ref{usual DB}) holds.
In this case, the steady state probabilities $P_i$ obey Boltzmann distribution, the currents
$J_i$ vanish and there is no average displacement of the motor. When $F$ and $\Delta \mu$ are
not simultaneously zero, the detailed balance condition (\ref{usual DB}) is broken,
the system is out of equilibrium and currents are present.

\subsection{A discrete ratchet model}

\begin{figure}[ht]
{\par \includegraphics[scale=0.5]{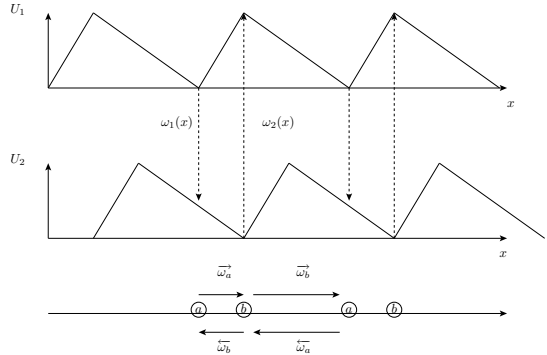}
\par}
\caption{\label{fig:ratchet} The top two curves represent the two time independent periodic potentials
$U_1(x)$ and $U_2(x)$ of the flashing ratchet model. At any position $x$, vertical transitions are possible between
the two internal states with rates $\omega_1(x)$ and $\omega_2(x)$.  Below is represented
the corresponding discrete model, which is obtained by considering effective transitions between
 the minimum of $U_1(x)$ (state a) to the minimum of the other potential (state b),
with rates as shown in the lower part of the figure. }
\end{figure}

From the  continuous model, a simplified effective discrete model can be constructed
as shown schematically in figure \ref{fig:ratchet} following the procedure outlined in \cite{widom}.
\begin{figure}
{\par { \rotatebox{0}{\includegraphics[scale=0.5]{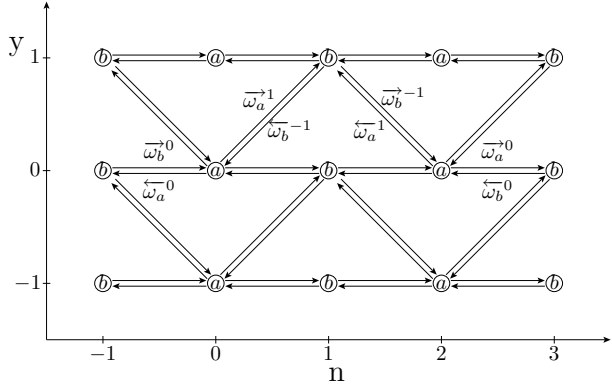}} }
\par}
\caption{A schematic of the evolution of the motor in a plane
$(n,y)$, where $n$ represents the position of the motor on the
filament and $y$ is the number of ATP molecules consumed. The even
and odd sublattices are denoted by $a$ and $b$, respectively.
Note that the lattices of $a$ and $b$ sites extend infinitely in
both directions along the $n$ and $y$ axis. The possible
transitions are represented with arrows on a particular section of
the lattice. } \label{fig:sketch}
\end{figure}
We assume that the motor has  a  vanishingly small
 residence time in all states which are not minima of the potentials $U_1(x)$ or $U_2(x)$, i.e.
 the time taken to slide down towards  a  well is negligible.
 Switching  between the potentials, represented as dashed lines in the figure, occurs at finite rates
$\omega_1(x)$ and $\omega_2(x)$, but only between states that are at the same location $x$.
  Since  downward sliding  occurs  instantly, the observable transitions are
effectively from one minimum of one potential (state a) to the other minimum of the other
potential (state b).
In this way, a discrete hopping model on a 1D lattice is constructed
in which transitions are allowed between even and odd sites called $a$ and $b$.
The  dynamics of the motor  on the   linear discrete lattice is as follows: the motor
 hops  from one site to neighboring sites, either consuming or producing ATP (see Fig.~\ref{fig:sketch}).
The position of the motor is denoted by $x = n d$, where $2 d \approx 8 \, \mbox{nm}$ is
 the step size of a kinesin. The even sites (denoted by $a$) are the low-energy state of the
motor, whereas the odd sites (denoted by $b$) are its high-energy state;
their energy difference is $ \Delta E \equiv k_B T \epsilon$, where $k_B$ is the Boltzmann constant and
$T$ is the temperature.  Because of the periodicity of the filament, all the
even ($a$) sites and all the odd ($b$) sites are equivalent. The dynamics of the motor is governed by
a master equation for the probability, $P_n(y,t)$, that the motor has consumed $y$ units of ATP and is at site $n$ at time $t$:
\begin{eqnarray}
\partial_t P_n(y,t)  = -  \left( \overleftarrow{\omega}_n + \overrightarrow{\omega}_n \right) P_{n}(y,t)
\phantom{\overleftarrow{\omega}_{n+1}^{\,l} P_{n+1}(y-l,t)}\nonumber \\
+  \sum_{l = -1,0,1} \left [\,
\overleftarrow{\omega}_{n+1}^{\,l}\,P_{n+1}(y-l,t)+
\overrightarrow{\omega}_{n-1}^{\,l}\,P_{n-1}(y-l,t) \,\right ],
\label{complete master eq}
\end{eqnarray}
where $\overleftarrow{\omega}_n \equiv \sum_l \overleftarrow{\omega}_{n}^{\,l}$ and
$\overrightarrow{\omega}_n \equiv \sum_l \overrightarrow{\omega}_{n}^{\,l}$.
Denoted by $\overleftarrow{\omega}_{n}^{\,l}$ and $\overrightarrow{\omega}_{n}^{\,l}$
are the transition rates  for the motor  to jump from site $n$ to $n-1$ or  to $n+1$, respectively,
 with $l\,(= -1,0,1)$ ATP molecules consumed.

As we show below, this discrete stochastic model contains the essential features of the original
ratchet model while being more amenable to precise mathematical
analysis \cite{kolomeisky,nelson,mazonka,lipowsky-2000}. In this sense, the discrete model
may be regarded as a \emph{minimal} ratchet model.

\subsection{Application of the model to experiments}

\subsubsection{Modes of operation and efficiency:}

\begin{figure}[ht]
\includegraphics[height=1.9in]{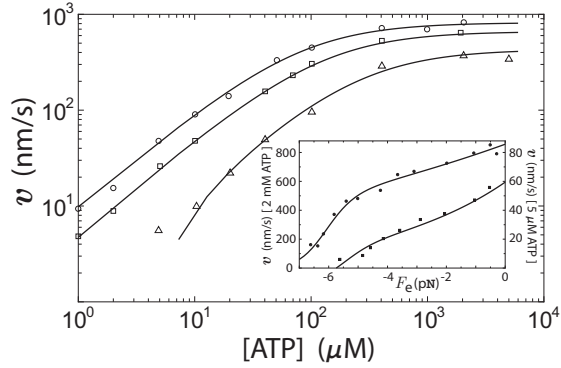}
\caption{Kinesin velocity vs.\ ATP concentration under an external
force \cite{prl-FT}. The solid curves are the fits of our model to
data from Ref.\ \cite{block}. From  top to  down, the plots are for
$F_e = -1.05, -3.59$, and $-5.63\,\mbox{pN}$, respectively. Inset:
Kinesin velocity vs.\ force under a fixed ATP concentration.  The
solid curves are fits to the data of Ref.\ \cite{block}. From
top to  down, the plots are for $[\mbox{ATP}] = 2\,\mbox{mM}$ and
$5\,\mbox{$\mu$M}$.} \label{fig:fits}
\end{figure}
\begin{figure}
\includegraphics[height=1.9in]{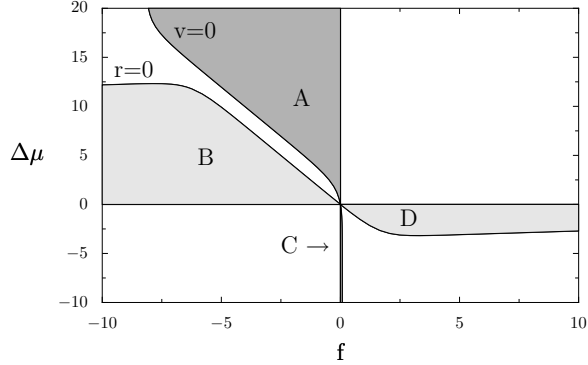}
\caption{The four modes of operation of a molecular motor (such as kinesin)
are delimited by $\bar{v}=0$ and $r=0$. The lines are
generated with parameters that we have extracted by  fitting the
data for kinesin in Ref.~\cite{block} to our model, and this fit
is shown in fig.~\ref{fig:fits}. } \label{fig:modes_operation}
\end{figure}

From the master equation Eq.~\ref{complete master eq}, one can
obtain the average velocity of the motor $\bar{v}$ and its average
ATP consumption rate, $\bar{r}$. One finds explicitly that
 \be \bar{v}=2 \frac{\Wr_a \Wr_b - \Wl_a \Wl_b}{\Wr_a+\Wr_b+\Wl_a+\Wl_b}
\label{explicit v} \ee \be r= \frac{\left( \Wl_a^1+\Wr_a^1 \right)
\left( \Wr_b+\Wl_b \right) - \left( \Wl_b^{-1}+\Wr_b^{-1} \right)
\left( \Wr_a+\Wl_a \right)} {\Wr_a+\Wr_b+\Wl_a+\Wl_b} \, .
\label{explicit r} \ee

By modelling the dependence of the rates on the force and on the
chemical potential in a way similar to what was done in
Eq.~(\ref{rates omega}) for the flashing ratchet model, one
obtains a theoretical prediction for the dependence of the
velocity and average ATP consumption rate on the force or the ATP
concentration, that can be compared to experiments. Despite its
simplicity, this discrete model can account quantitatively for
such measurements as shown in figure (\ref{fig:fits}) in the case
of a kinesin.

From the two currents $\bar{v}$ and $r$, a diagram of operation of
the motor (see  figure \ref{fig:modes_operation}) can be
constructed, which summarizes the possible thermodynamic modes of
operation of the motor \cite{prl-FT}. This diagram is similar to
that given in Ref.~\cite{armand1}, except that the present diagram
extends to the regime far from equilibrium rather than being
limited to the linear response regime. Whenever, $f \bar{v}<0$
work is performed by the motor, whenever $r \Delta \mu<0$ chemical
energy is generated. The motor can work in eight different
regimes. Four of them are passive and correspond to the white
regions in Fig.~\ref{fig:modes_operation}, in which there is no
energy output from the system, since $f \bar{v}>0$ and $r \Delta
\mu>0$.
\footnote{The case where $f \bar{v}<0$ and $r \Delta \mu<0$ is
forbidden by the second law of thermodynamics, because it would lead
to a negative entropy production. For this reason, there is no point
in Fig.~\ref{fig:modes_operation} which corresponds to this case.}
The four remaining
regimes are more interesting since $f \bar{v}$ and $r \Delta \mu$
are not of the same sign, which means that some form of
transduction occurs between the mechanical and chemical forms of
energy. More precisely:

\begin{itemize}
\item
In Region A of the diagram, where $r \Delta \mu
>0$ and $f \bar{v}<0$, the motor uses the chemical energy of ATP to
perform mechanical work. This can be understood by considering a
point on the y-axis of Fig.\ \ref{fig:modes_operation} with
$\Delta \mu>0$. There we expect that the motor drifts to the right
with $\bar{v}>0$. Now in the presence of a small load $f<0$ on the
motor, we expect that the motor is still going in the same
direction although the drift is uphill and thus work is performed
by the motor at a rate $\dot{W}=-f\bar{v} >0$. This holds as long
as $f$ is smaller than the stalling force, which defines the other
boundary of region A.

\item
Similarly, in Region B, where $r \Delta \mu
<0$ and $f \bar{v}>0$, the motor produces ATP already in excess from mechanical
work.

\item
In Region C, where $r \Delta \mu >0$ and $f \bar{v} <0$, the
motor uses ADP in excess to perform mechanical work.

\item
In Region D, where $r
\Delta \mu <0$ and $f \bar{v}>0$, the motor produces ADP already in excess from
mechanical work.
\end{itemize}
It is interesting to note that the large
asymmetry between regions A and C in Fig.\
\ref{fig:modes_operation} reflects the fact that kinesin is a
unidirectional motor.
The diagram also illustrates the fact that under  usual conditions with
$\Delta \mu \simeq 10-25$, a kinesin uses the chemical energy of
ATP hydrolysis to produce mechanical work (region A of the
figure), rather than operating  in the other way to synthesize ATP
(region B of the figure).

\begin{figure}[ht]
\includegraphics[height=1.9in]{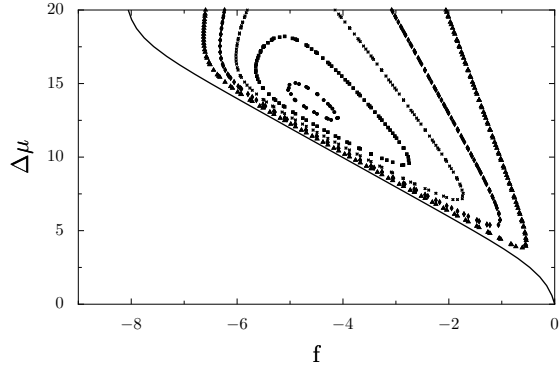}
\caption{Curves of equal efficiency $\eta$ within region A (which
is delimited by the solid line and by the $y$ axis). The
parameters are those used in fig.~\ref{fig:modes_operation} and
obtained from the fit of
 fig.~\ref{fig:fits}. From the outside to the inside the curves
correspond to   $\eta=0.2$,
  $\eta=0.3$,   $\eta=0.4$, $\eta=0.5$ and $\eta=0.58$.
 The absolute maximum efficiency for these
parameters is about 59\% and is located at $\Delta \mu \simeq 14$
and $f \simeq -4.9$. } \label{fig:efficacité}
\end{figure}
 It is also possible to analyze  the mechanical efficiency of this motor,
defined as the ratio between the mechanical work delivered by the
motor divided by the chemical energy supplied \cite{pre-FT}. Other
definitions of efficiency have been considered in the literature (such
as the efficiency at maximum power \cite{seifert-epl-2008} 
or the Stokes efficiency) but the advantage of this definition is that it holds arbitrary far from
equilibrium and it corresponds near to equilibrium to the definition used
traditionally with heat engines. The kinesin operates most
efficiently in a range of values of $\Delta \tilde{\mu}$ which
corresponds well to the typical free energy delivered by the
reaction of ATP hydrolysis (physiological conditions correspond to
$\Delta \mu \simeq 10-25$). The maximum of efficiency is obtained
around a single isolated point in the coordinates $(f,\Delta \mu)$
(rather than on a line as in the near
 equilibrium regime for instance) as shown in figure \ref{fig:efficacité}.
 This suggests that kinesin is in fact optimized to operate
 under a load corresponding to a normalized force of about $-4.9$
 in the conditions of Fig.~\ref{fig:efficacité}.
The maximum of efficiency is around $40-60 \%$, much higher than
the typical efficiency in the near equilibrium regime (of the
order of $0.03 \%$). The value of the maximum efficiency of $40-60
\%$ agrees well with recent measurements for kinesin.

\subsubsection{Violation of Onsager and Einstein relations:}

\begin{figure}
\includegraphics[height=1.9in]{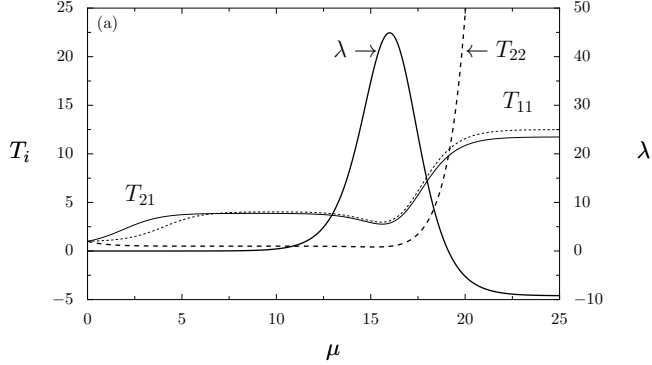}
\caption{Effective temperatures $T_{11}$ (dot-dashed), $T_{21}$
(dotted), $T_{22}$ (dashed), and $\Delta \lambda$ (solid) vs.\
$\Delta \mu$ in Region A of Fig.\ \ref{fig:modes_operation} for a
small $f$. Note that $T_{ij}$ characterizes the
fluctuation-response ratios (see text), while $\Delta \lambda$
quantifies the breaking of Onsager symmetry.}
\label{fig:effect-temp}
\end{figure}
Away from equilibrium, we expect that Onsager and Einstein
relations are no longer valid. To quantify their violations, we
have introduced in Ref.~\cite{prl-FT} $\Delta \lambda \equiv
\lambda_{12}- \lambda_{21}$ to quantify the violation of Onsager
relations and four ``temperature"-like quantities, $T_{ij} \equiv
D_{ij}/\lambda_{ij}$ to quantify the violation of Einstein
relations. All these quantities are defined using linear response
theory in the vicinity of  a non-equilibrium steady
 state rather than near  an
equilibrium state. Of course, these effective temperatures are not
thermodynamic temperatures: they are merely one of many possible
ways to quantify deviations of Einstein relations. These $T_{ij}$
and $\Delta \lambda$ are shown in figure \ref{fig:effect-temp} as
functions  of $\Delta \mu$ for the particular case of $f \ll 1$
within region A. We observe that all the $T_{ij}$ start off at
$T_{ij}=1$ near equilibrium where $\Delta \lambda=0$ as expected
from Onsager relations, whereas for large $\Delta \mu$, $T_{22}$
diverges exponentially while $T_{21}$, $T_{11}$ and $\Delta
\lambda$ approach constant values  \cite{prl-FT}.

\subsubsection{Beyond two states model of molecular motors}

The discrete two states model presented above describes many
features of experiments on a single kinesin such as the average
velocity versus force or versus ATP concentration,  or the average
ATP consumption rate. It does not describe however equally well
the fluctuations of these quantities. As shown in
\cite{kolomeisky-revue}, at least four internal states are necessary
to describe the fluctuations of position of the motor, which are
quantified by the so called randomness parameter introduced and
measured in Ref.~\cite{block-1999}.
For this reason, a more refine model of kinesin should contain
more than two internal states to account for the way the motor
walks on the filament, which is by a succession of binding and
unbinding events of the two heads in a hand-over-hand fashion. To
include that aspect, a model with 9 states - which can be reduced
to 7 states for most cases - was proposed in
Refs.~\cite{lipowsky:2008,lipowsky-valleriani:08,lipowsky:2009},
where the 7 states describe the most significant chemical states
of the two headed kinesin. The possible transitions between these
states can be represented by a network, which describes the
mechano-chemical coupling in this motor. In this network
representation, several cycles can be identified just like in the
discrete model presented above. The model successfully accounts
for many experimental results known for kinesins \cite{carter}.
Using this framework, a diagram summarizing the thermodynamic
modes of operation of the motor has been constructed in
Ref.~\cite{liepelt:2009}. This diagram has similarities with our
figure \ref{fig:modes_operation}, but some differences are present
due to the different role played by the mechanical and chemical
cycles in the different models.

Many other works have used discrete or continuous
stochastic models to analyze molecular
motors : for instance a discrete model with 7 states
has been developed for myosin V \cite{mturner:2006}.
A discrete model with only 3 states has been used to describe the fluctuations
of position of nucleosomes along DNA in Ref.~\cite{Mollazadeh-Beidokhti2009}.
In Ref.~\cite{gaspard-gerritsma}, the rotating motor $F_1$-ATPase is described
by a stochastic process for the angle of rotation of the motor, which is treated as a continuous variable and for the chemical states, which are treated as discrete states. More recently, the authors of this reference have
 developed a discrete version corresponding to their continuous model, which is a
 two states model very similar to the one discussed in this review \cite{gaspard-gerritsma2}.

\section{Fluctuation relations in models of molecular motors}
Fluctuation relations quantify the exchanges of energy between a
system and its environment when the system is in a non-equilibrium
state \cite{kurchan,ritort-2008}. These relations hold arbitrarily
far from equilibrium in a regime where the usual thermodynamic
laws - which hold only near equilibrium - do not apply. Since
their discovery about a decade ago
\cite{GC:1995,Jarzynski:1997,kurchan,evans,crooks}, there has been
a growing interest to understand their importance and
implications. One reason for the popularity of this topic has to
do with the fact that these relations provide a fresh look at old
fundamental questions, such as the origin of irreversibility or
the second law of thermodynamics.

For small systems (for which the fluctuations are large, in the
sense that their magnitude can be of the same order as  the
average value), the fluctuation relations impose new constraints
which go beyond the usual description of statistical fluctuations.
Many fluctuation relations have been verified experimentally using
biopolymers, in particular the Jarzynski's relation
\cite{Liphard-2002}, the Crooks relation \cite{collin-2005} and
the Hatano-Sasa relation \cite{ritort-sasa}. Complementary
experimental verifications have been carried out with colloidal
particles in optical traps \cite{imparato:2007,blickle-2006}.
Recently, a modified Fluctuation-Dissipation relation, related to
the Hatano-Sasa relation has been verified for a colloidal
particle in a nonequilibrium steady state \cite{chetrite-2009}.
All these experiments represent remarkable achievements, which
confirm the validity of the general framework of fluctuation
relations in various experimental conditions. However, it may be
worth pointing out that in all these examples, the experiments
have been designed in order to verify the fluctuation relations.
On the contrary, the case of molecular motors is particularly
interesting since this is a system which was not designed for that
objective. Molecular motors operate in a regime far from
equilibrium, with fluctuation relations in some sense built-in in
their natural mode of operation.
For this reason, it is more appropriate to think about the
fluctuations relations as thermodynamic constraints on the
operation of the motors, which presumably theoretical models of
molecular motors should obey \cite{gaspard2,seifert}. That of
course would assume that the fluctuations relations are obeyed
exactly in experiments. To our knowledge, quantitative
experimental tests of fluctuation relations have not been carried
out on molecular motors yet, although there is some indication
that,  for instance,  the data of \cite{carter} on single molecule
experiments with kinesin is in agreement with the fluctuations
relations.

The dynamics of a molecular motor breaks the detailed balance condition,
and leads to a non-equilibrium steady state characterized by the presence
of non-zero currents, which are independent of time. For each current, one
can associate a cycle, also called an irreversible loop. The construction
of these cycles and the way they can be associated with  currents is explained
by a general theory for systems governed by  a master equation \cite{schnakenberg,hill-book1,gaspard-2007}.
One central result of this theory is the following relation
\begin{equation}
\frac{ \Pi^+ ({\mathcal L})}{ \Pi^- ({\mathcal L})}=\frac{J^+}{J^-}=e^{A/{k_B T}},
\label{dedonder}
\end{equation} where $\Pi^+({\mathcal L})$ denotes the
product of reaction rates associated with  the different transitions within
the cycle ${\mathcal L}$ in the clockwise direction, whereas
$\Pi^-({\mathcal L})$ denotes the same product in the counter clockwise direction. We
have denoted $J^+$ the number of cycles undergone by the motor  per unit time
in the clockwise direction, and $J^-$ the number of cycles undergone per unit time in
the counter clockwise direction, so that overall the cycle flux is $J=J^+ - J^-$.
The quantity $A$ is called affinity or thermodynamic force. This affinity
can be defined as the derivative of an effective potential, experienced by a biased random walker that would exhibit the same dynamics \cite{pre-FT}.
When the detailed balance condition is satisfied, $\Pi^+({\mathcal L})=\Pi^-({\mathcal L})$, the effective
potential is flat and $A=J=0$.
\begin{figure}[ht]
\begin{center}
\includegraphics[scale=0.8]{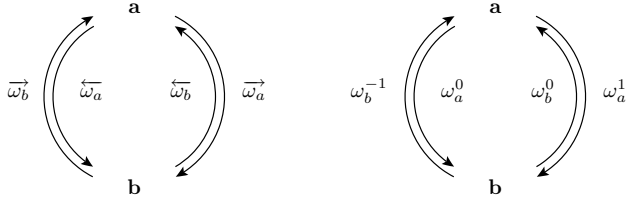}
\end{center}
\caption{Cycles associated with the evolution of the motor in the discrete two-states model.
Left: the cycle for the position variable $n$; the  length $n$ run by the motor corresponds to
half the number of turns run around the cycle (the factor 1/2 has to do with the period of the motor
which is twice the  unit length of the lattice on which the motor evolves). Right:
the cycle for the chemical variable $y$. } \label{fig:representation_cycles}
\end{figure}

In the simple case where the model contains a single cycle with
only two states as in
 figure \ref{fig:representation_cycles}, the
relation (\ref{dedonder}) leads to the affinity $A/k_B T= -2 \Psi$
where $\Psi$ is defined by
\begin{equation}\label{Effective potential}
\Psi= \frac{1}{2} \ln \left( \frac{\Wl_a \Wl_b}{\Wr_a \Wr_b}
\right),
\end{equation}
and the corresponding current $J$ is  the average motor velocity
 (see Eq.~\ref{explicit v}):  \be
\bar{v}=2 \frac{\Wr_a \Wr_b - \Wl_a
\Wl_b}{\Wr_a+\Wr_b+\Wl_a+\Wl_b}.  \ee

In order to describe more precisely the dynamics of this system,
let us consider $P_i(n,t)$, the probability that the motor at time
$t$ is  on the site $i$ $(=a,b)$ and at the position $n$ (with
$x=nd$ where $d$ is distance between sites $a$ and $b$). This
probability can be obtained for instance by integrating over the
variable $y$ in the quantity $P_n(y,t)$, which satisfies the more
general master equation of Eq.~(\ref{complete master eq}). Because
of the periodicity of this problem, it is convenient to introduce
generating functions $F_i(\lambda,t) \equiv \sum_{n} e^{-\lambda n
} P_{i}(n,t),$ which evolve according to : $\partial_t F_i=
{\mathcal M}_{ij}\,F_j$, where ${\mathcal M}[\lambda]$ is the
following $2\times 2$ matrix constructed from the master equation
satisfied by $P_i(n,t)$:
\begin{eqnarray} {\mathcal M}[\lambda] =\left [
 \begin{array}{cc}
  -\overrightarrow{\omega_a}-\overleftarrow{\omega_a} &
  e^{\lambda}\, \overleftarrow{\omega_b} + e^{-\lambda}\,
  \overrightarrow{\omega_b} \\ e^{\lambda}\,
  \overleftarrow{\omega_a}
  +  e^{-\lambda}\,\overrightarrow{\omega_a}
   & -\overleftarrow{\omega_b}-\overrightarrow{\omega_b} \\
\end{array} \right ]. \nonumber
\end{eqnarray}
In the long time limit, the steady state properties of the motor can
be obtained from the largest eigenvalue $\vartheta[\lambda]$
of this matrix. Indeed when $t \rightarrow \infty$,
  \be \left \langle\,e^{-\lambda n }\,\right
\rangle = \sum_{i} F_i(\lambda,t) \sim \exp \left( \vartheta\,t
\right). \label{Evolution for F} \ee The first derivative of
$\vartheta$ with respect to $\lambda$ gives the average
velocity $\bar{v}$  of the motor and  the second derivative  gives
the diffusion coefficient of the motor.

From the explicit expression of this eigenvalue, the following property
may be derived
\begin{equation}\label{GC for theta} \vartheta(\lambda)=
\vartheta( -\Psi- \lambda),
\end{equation} which is the Gallavotti-Cohen  fluctuation theorem. Other equivalent forms of this relation can be
obtained. One of them involves the large deviation function of the current $v$ denoted $G(v)$,
defined in the long time limit,  as:
\begin{equation}\label{LD_def}
P(\frac{n}{t}=v) \sim e^{-G(v)t} \, .
\end{equation}
The  analytical expression of this function,
  obtained  in Ref.~\cite{pre-FT},  has a complicated non-linear expression in terms of the rates, but it
satisfies a surprisingly simple relation:
\begin{equation}\label{GC for LD}
G(v)-G(-v)=  \Psi v.
\end{equation}
This relation implies that the ratio of the probabilities to observe a velocity
$v$ or $-v$ after a time $t$ satisfies the
relation:
\begin{equation}\label{GC for P}
\frac{\mathcal{P}(\frac{n}{t}=v)}{\mathcal{P}(\frac{n}{t}=-v)}=e^{-
\Psi vt}.
\end{equation}
Using Eq.~(\ref{GC for LD}), and the fact that $G(v)$ and
$\vartheta(\lambda)$ are Legendre transforms of each other,
one recovers indeed the relation~(\ref{GC for theta}).

The relations~(\ref{GC for theta}),(\ref{GC for LD}) and (\ref{GC for P})
are equivalent forms of a constraint imposed on the system by the
Gallavotti-Cohen fluctuation theorem. This theorem
itself is a consequence of the time-reversal symmetry of the  physical
 laws involved in this model. This
 symmetry is a  fundamental property  that does not depend
 into  on the details of  the system
 and therefore, in this sense, the  fluctuation theorem appears as  a universal
 requirement, just as thermodynamic constraints are universal for
 systems at equilibrium regardless of their microscopic structure.
   Of course,  universality does not imply that
  the constraints are
  easy to find and to formulate  explicitly  for a given  problem.
  Here, the relation~(\ref{GC for P})  is  an  explicit
  prediction derived from  the  fluctuation theorem for molecular motors.
  It would
  be of great interest to verify this relation experimentally. Conversely,
  this equation can be used to measure the affinity $\Psi$,  and therefore
  to access indirectly  the microscopic rates,  for a given molecular motor.

\subsection{Modeling processivity at the single motor level}
Experiments on single molecular motors depend on an important
property of these motors called processivity. Molecular motors
like kinesins, which can hydrolyze a large number of ATP molecules
before detaching from microtubules, are called processive, whereas
those like myosins II, which detach and reattach frequently from
actin filaments and are called non-processive. There are several
ways to define processivity, either it can be defined as the
average lifetime of the motor on the filament, or from the average
length spanned by the motor or from the average number of ATP
molecules consumed before detaching. In single molecule
experiments \cite{schnitzer:00,carter}, the dependence of the
run-length of a single kinesin on load and on the ATP
concentration has been studied. On the theoretical side, the
average lifetime of a molecular motor as function of load has been
studied in \cite{peliti-prost2001} using the flashing ratchet
model. Here we focus on the run-length, for which we derive a
simple expression within the discrete two states model presented
above. Using a similar theoretical approach, the average time
before observing a backward step in a discrete model of a
molecular motor has been studied in Ref.~\cite{gaspard2}. More
generally, in a network of discrete chemical states, there are
well known methods to calculate the average lifetime of a random
walker in the presence of absorbing states \cite{hill-book1}.
These methods can be used not only to calculate the lifetime of
the motor on the filament, but also the dwell times associated
with the motor steps as shown in
Ref.~\cite{lipowsky-valleriani:08}.

The detachment of the motor from the filament can be represented
by a detachment rate $\kappa$, which depends on the local site on
the filament visited by the motor. Since in the discrete model
presented above, the state $b$ is the high energy state and $a$ a
low energy state, we assume for simplicity that detachment only
occurs from site $b$, which corresponds to the maxima of $U_1(x)$
in figure \ref{fig:ratchet}. Because of this detachment, the motor
can be no longer only in states $a$ or $b$, thus we need to modify
the master equation in order to conserve probability at all times.
This can be done by adding an extra state corresponding to the
unbound motor state, which is an absorbing state. We define the
generating functions $F_i(\lambda,t)$ as before, but now the
matrix of evolution of these generating function is the following
3x3 matrix:
\begin{eqnarray} {\mathcal M}[\lambda] =\left [
 \begin{array}{ccc}
  -\overrightarrow{\omega_a}-\overleftarrow{\omega_a} &
  e^{\lambda}\, \overleftarrow{\omega_b} + e^{-\lambda}\,
  \overrightarrow{\omega_b} & 0 \\ e^{\lambda}\,
  \overleftarrow{\omega_a}
  +  e^{-\lambda}\,\overrightarrow{\omega_a}
   & -\overleftarrow{\omega_b}-\overrightarrow{\omega_b}-\kappa & 0 \\
   0 & \kappa & 0 \\
\end{array} \right ]. \nonumber
\end{eqnarray}

This matrix has three eigenvalues $\mu_1$, $\mu_2$ and 0, which is associated with the absorbing state.
The corresponding eigenvectors are $| \mu_1 \rangle$, $| \mu_2 \rangle$ and $| c \rangle$. If the initial
state vector is $| F(\lambda,0) \rangle = A | \mu_1 \rangle + B | \mu_2 \rangle + C | c \rangle$, the state
 vector at time $t$ is $| F(\lambda,t) \rangle = A e^{\mu_1 t} | \mu_1 \rangle + B e^{\mu_2 t}
| \mu_2 \rangle + C | c \rangle$. Since $\mu_1$ and $\mu_2$ are
strictly negative, at time $t \rightarrow \infty$, $\langle
e^{-\lambda n} \rangle = \langle 0 | F(\lambda, \infty)
\rangle=C(\lambda)$, with $\langle 0|=(1,1,1)$. Note that
$C(\lambda)$ contains all the moments of the run length, and in
particular the average run length, which is, in units of the
lattice period, \be \langle n \rangle = - \frac{\partial
C(\lambda)}{\partial \lambda}|_{\lambda=0}. \ee The function
$C(\lambda)$ can be calculated by projecting the left eigenvector
associated the eigenvalue 0, $\langle c|$, onto the initial state
vector  $| F(\lambda,0) \rangle$. If the initial state vector is
along the unit vector ${\bf e}_x$, then one obtains
\begin{equation}\label{C_lambda}
C(\lambda)=\frac{\Wr_a e^\lambda + \Wl_a e^{-\lambda}}{r \omega_a + \Wr_a \Wr_b (1 - e^{2 \lambda} )
+ \Wl_a \Wl_b (1- e^{-2 \lambda})}.
\end{equation}

If the initial state vector is along the unit vector ${\bf e}_x$,
then \be \langle n \rangle= \frac{\Wr_a - \Wl_a}{\omega_a +
\omega_b} + 2 \frac{\Wr_a \Wr_b - \Wl_a \Wl_b}{\kappa \omega_a}.
\ee If the initial state vector is along the unit vector ${\bf
e}_y$, then \be \langle n \rangle=  2 \frac{\Wr_a \Wr_b - \Wl_a
\Wl_b}{\kappa \omega_a}. \ee Thus the part of the average length
which is independent on the initial condition is \be  \langle
\bar{n} \rangle= 2 \frac{\Wr_a \Wr_b - \Wl_a \Wl_b}{\kappa
\omega_a} = \frac{\bar{v}}{\kappa P_b}, \ee where $\bar{v}$ is the
average motor velocity defined before and $P_b$ the stationary
probability to be in state $b$ when $\kappa=0$.

Note that it has been assumed implicitly that the motor runs in
the positive direction so that by construction $\bar{v}>0$,
$\langle \bar{n} \rangle$ is positive and has the familiar form
obtained above. The application of fluctuation relations to
non-processive motors has not been discussed in the literature to
our knowledge. We believe that a fluctuation relation will be
obeyed only if a reattachment process is taken into account in the
model. In this case, there is no absorbing state anymore.

\subsection{Mechanochemical coupling for the discrete model}
We have so far only discussed the form of fluctuation relations
for models containing a single cycle. It is well known that models
with at least two cycles must be introduced to account for
experimental data such as those represented in Figure
\ref{fig:fits}. In order to discuss more general fluctuation
relations, and to compute in a simple way the chemical current,
$r$, associated with the average ATP consumption rate, it is
necessary to include in the description of the state of the motor,
a chemical variable $y$ associated with the average number of ATP
consumed as done in Eq.~(\ref{complete master eq}). With the
notations, $\omega_a^l=\Wr_a^l+\Wl_a^l$ and $\omega_a=\Wr_a+\Wl_a$
(and similarly for site $b$), one obtains the chemical current
 \be r= \frac{\omega_a^1 \omega_b -
\omega_b^{-1} \omega_a} {\omega_a + \omega_b}, \,  \ee
 in agreement with the cycle representation of figure
\ref{fig:representation_cycles} and with
 the formula~(\ref{explicit r}) above.

When the form of these rates is explicitly given in terms of the
normalized force $f$ and the normalized chemical potential $\Delta \mu$, a new reformulation
of Eq.~(\ref{dedonder}) is obtained:
\begin{equation}\label{Steady state balance condition}
k_B T \ln\frac{\overrightarrow{\omega_b}^{-l}
\overrightarrow{\omega_a}^{l'}}{\overleftarrow{\omega_a}^{l}
\overleftarrow{\omega_b}^{-l'}}= F_e (2d) - \Delta \tilde{\mu}
\left( l-l' \right),
\end{equation}
with $l,l'=0,1$. This equation can be understood as a statement of
the first law of thermodynamics at the level of elementary
transitions \cite{lipowsky:2008,schmiedl-2007}. Indeed, it is
possible to associate the left hand side of this equation with the
heat released by the motor into the environment (treated as a
reservoir at the same temperature) during transitions $(l,l')$.
The right hand side can then be interpreted as the difference
between the mechanical work $-F_e (2d)$ and the variation of
chemical energy $\Delta \tilde{\mu} \left( l-l' \right)$ for these
transitions. The variation of internal energy is zero in this case
since only transitions involved in a cycle are considered. Note
that similarly to this, the generalized detailed balance condition
of Eq.~(\ref{generalized DB}) can be also interpreted as a
statement of the first law at the level of elementary transitions.

Following the same steps that lead to the fluctuation relations
for models with one cycle but now for a model with two cycles, one
arrives at the following relation, similar to
 Eq. ~(\ref{GC for theta}):
\begin{equation}\label{GC for theta meca-chimie}
\vartheta(\lambda,\gamma)= \vartheta(-\tilde{\Psi} - \lambda,
-\tilde{\chi}- \gamma),
\end{equation}
with new affinities $-\tilde{\Psi}$ and $-\tilde{\chi}$
associated with  the mechanical and chemical cycles. These affinities
represent a part of the expression of the entropy production rate of the
motor \cite{pre-FT}, which also satisfies a fluctuation relation different from that
of the currents but very much related to it.

 We emphasize that the  relation~(\ref{GC for theta meca-chimie})
 involves {\it both} the mechanical and the chemical activities
 and that, here, a  symmetry relation of the type of Eq.~(\ref{GC for theta}) for the
 mechanical variable alone is  not  satisfied. It is therefore essential, in deriving
 fluctuation relations, to take into account and include all internal variables that are coupled
 with one another  under time-reversal. Leaving aside some relevant degrees of freedom
 would manifest itself as an apparent  violation of this symmetry and  would  wrongly
 be interpreted as   a breakdown of the fluctuation theorem.

\subsection{Flashing ratchet model on a continuous space}
\begin{figure}[ht]
\begin{center}
\includegraphics[scale=0.8]{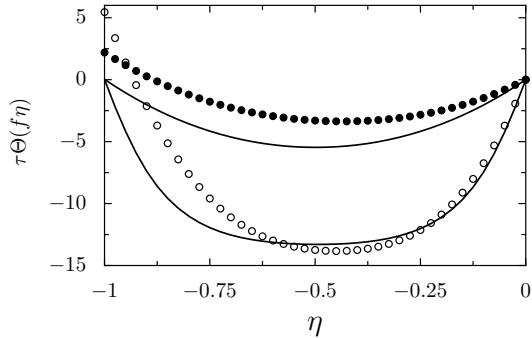}
\end{center}
\caption{Normalized eigenvalue $\tau \Theta(f \eta)$ of the
flashing ratchet model as a function of $\eta=\lambda/f$ (with $\tau=a^2/D_0$),
for a normalized force $f=5$ (top two curves) and $f=10$ (bottom two curves).
The solid curves correspond to the case where the transition rates between the internal
states satisfy detailed balance, which leads to the Gallavotti-Cohen symmetry, {\it i.e.} to
the symmetry with respect to $\eta=-1/2$. The curves with filled symbols
($f=5$) and empty symbols ($f=10$) correspond to the case where the detailed balance is broken with constant
transition rates $\omega_1(x)=\omega_2(x)=10 \tau^{-1}$ and the same potentials.}
\label{fig eigenvalue_switching}
\end{figure}

We show in this section how the ideas developed in the previous section for the discrete two
states model can now be extended to more general continuous models, such as the
flashing ratchet.
There are several reasons for considering
continuous models as a substitute for discrete models: first of all, continuous models
 contain all the possible discrete models as limiting cases, second, the way to describe
the effect of force on the motor is unambiguous for continuous models, and third,
there are effects such as fluctuations which are not always well captured by discrete models.

We provide in this section  an analytical proof that the flashing
ratchet obeys a  Gallavotti-Cohen symmetry
\cite{flashing-ratchet},  using a technique
inspired by \cite{kurchan,lebowitz}. We also analyze numerically
this point by calculating the eigenvalue associated with the
evolution matrix of the generating functions of the currents.

\subsubsection{The purely mechanical ratchet}
Before considering the case of the flashing ratchet with a
mechanical and a chemical variable, it is helpful to look first at
a purely mechanical ratchet, which has been used to describe in
particular the translocation of a polymer through a pore
\cite{dlubensky}. In this model, one considers a random walker in
a periodic potential subject to an external force $F$ (model I)
\cite{risken,hanggi}. The corresponding Fokker-Planck equation is
\begin{equation}
   \frac{\partial P}{\partial t} = D_0 \frac{\partial}{\partial x} \left[
  \frac{\partial P}{\partial x}  +
 \frac{{U'}(x) - F }{k_B T} P \right],
 \label{eq:FPuncoupled}
\end{equation}
where $U(x)$ is a periodic potential $U(x + a) =  U(x)$ and $a$ is the period.
This equation describes the stochastic dynamics of a particle
 in the effective potential $U_{eff}(x)
  = U(x) - Fx$.
  By solving Eq.~(\ref{eq:FPuncoupled}) with periodic
boundary conditions \cite{dlubensky,risken}, it can be readily proven that
  the system reaches a stationary state with
 a uniform current $J$ in the long time limit. This current is
 non-vanishing if a non zero force is applied. When $F=0$, there is no tilt
in the potential, $J=0$ and the stationary probability is given
by the equilibrium Boltzmann-Gibbs factor.

Similarly to the discrete case, we  introduce the generating function
  \begin{equation}
   F_{\lambda}(\zeta, t) = \sum_n \exp\left( \lambda(\zeta +n) \right)
 P( (n+\zeta)a, t)   \, .
\label{GenFunct}
 \end{equation}
The time evolution of this
generating function $F_{\lambda}$ is obtained by
summing over Eq.~(\ref{eq:FPuncoupled}). This leads to the following equation:
 \begin{equation}
  \frac{\partial F_{\lambda}(\zeta, t)} {\partial t }
 =   {\mathcal L}(\lambda)  F_{\lambda}(\zeta, t)   \, ,
\label{EvolGenFunct}
 \end{equation}
 where the  deformed differential operator ${\mathcal L}(\lambda)$ acts
 on a periodic function $\Phi(\zeta,t)$ of period $1$
 as follows:
  \begin{equation}
 \frac{a^2}{D_0} {\mathcal L}(\lambda)  \Phi =
 \frac{\partial^2 \Phi } {\partial \zeta^2 } +
 \frac{\partial }{\partial \zeta}  \Big(
   \tilde{U}'_{eff} \Phi  \Big)
 - 2 \lambda  \frac{\partial \Phi } {\partial \zeta  }
   -  \lambda \tilde{U}'_{eff}  \Phi + \lambda^2 \Phi \, ,
\label{def:Ldeformed}
  \end{equation}
where $\tilde{U}'_{eff}=a \partial_x U_{eff}/k_B T$ and
the left hand side of Eq.~(\ref{def:Ldeformed}) is proportional to
the inverse of the characteristic time $\tau=a^2/D_0$.

 The operator ${\mathcal  L}(\lambda)$  has the following
 fundamental  conjugation property:
\begin{equation}
  \hbox{e}^{U(x)/k_B T}
   {\mathcal  L}(\lambda)  \left( \hbox{e}^{-U(x)/k_B T}
   \Phi \right) =
  {\mathcal L^\dagger} \left(-f -\lambda \right)  \Phi.
 \label{OperatorGC1}
  \end{equation}
This property implies that the operators  ${\mathcal L}(\lambda)$
 and  ${\mathcal L^\dagger} \left(-f -\lambda \right)$ are
 adjoint to each other, and thus have the same spectrum.
If we call $\Theta(\lambda)$ the largest eigenvalue of
${\mathcal L}(\lambda)$,
we obtain from Eq.~(\ref{OperatorGC1})
that $\Theta(\lambda)$ satisfies the Gallavotti-Cohen symmetry:
\begin{equation}
 \Theta(\lambda) =  \Theta(-f -\lambda).
\label{GC1}
 \end{equation}
In fact, this symmetry holds for all eigenvalues.
 For the special  case $f=0$, the conjugation relation~(\ref{OperatorGC1})
 reduces to the {\it detailed balance} property \cite{lebowitz}.
One can note that this proof of the Gallavotti-Cohen symmetry does
not require explicit knowledge of $\Theta(\lambda)$. In the
discrete minimal ratchet model, an explicit analytical expression
could be obtained for this quantity. In the continuous case, this
is no longer the case but $\Theta(\lambda)$ can be calculated
numerically. We have done this by first discretizing
the operator ${\mathcal L}(\lambda)$ and then calculating its
largest eigenvalue using the Ritz variational method
\cite{flashing-ratchet}. A similar method has been used in
Ref.~\cite{mehl-seifert08} for the cosine potential. We note that
our numerical approach can handle any form of potential.

\subsubsection{The flashing ratchet}
We now present the extension of the Gallavotti-Cohen symmetry to
the case of the flashing ratchet model, which should include both
the mechanical and chemical currents
\cite{prl-FT,schmiedl-2007}.
 When the switching rates satisfy a detailed
balance condition, which is for instance the case when $\Delta
\mu=0$, the symmetry is indeed present as shown in the solid
curves of Fig.~\ref{fig eigenvalue_switching}. In the general case
however, where the normalized force $f$ and chemical potential
$\Delta \mu$ are both non-zero,
 the relation~(\ref{usual DB}) is no longer satisfied and
 the  Gallavotti-Cohen relation (\ref{GC1}) is not valid.
This is shown in the curves with symbols in Fig.~\ref{fig
eigenvalue_switching} where for simplicity we took constant
switching rates  $\omega_1=\omega_2=10 \tau^{-1}$. For all the
curves of this figure, a sawtooth potential $U_1$, and a potential
$U_2$ constant in space have been chosen. The breaking of the
symmetry of Eq.~(\ref{GC1})  can be interpreted as a result of the
existence of internal degrees of freedom. Although other
mechanisms exist which lead to violations of fluctuations
relations as discussed in Ref.~\cite{jarzynski_coarse_graining},
this case appears to be  rather generic.

\begin{figure}[ht]
\includegraphics[scale=0.8]{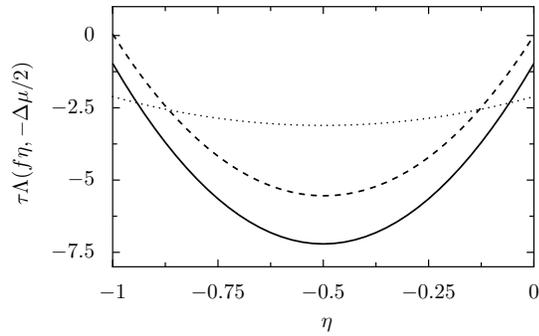}
\caption{For the  model described by
Eqs.~(\ref{eqs:moteur+chimie1}-\ref{eqs:moteur+chimie}), the normalized eigenvalue $\tau \Lambda(f
\eta,-\Delta \mu/2)$ is shown as function of $\eta$. The dashed
curve corresponds to $f=5$ and $\Delta \mu=0$,  the solid curve
corresponds to $f=5$ and $\Delta \mu=10$, and the dotted curve
corresponds to $f=2$ and $\Delta \mu=10$. The symmetry is
recovered in all cases in this description which includes both the
mechanical and chemical degrees of freedom.} \label{fig eigenvalue
2 variables}
\end{figure}

Let us now  introduce the probability density $P_i(x,q;t)$ associated
with the  probability that at time $t$
 the ratchet is in the internal state $i$, at position $x$ and
 that $q$ chemical units of ATP have been consumed.
 The evolution equations for this probability density is obtained
 by  modifying Eqs.~(\ref{eqs:moteur1}) after taking into
 account the dynamics of the discrete variable $q$. We have
 \begin{eqnarray}
 \frac{\partial P_1(x,q,t)}{\partial t} &=&
 \left( {\mathcal L_1}  - \omega_1(x) \right) P_1(x,q,t) \label{eqs:moteur+chimie1} \\
&+&  \omega_2^{-1}(x) P_2(x,q+1,t) +  \omega_2^{0}(x) P_2(x,q,t)
  \nonumber  \\
 \frac{\partial P_2(x,q,t)}{\partial t} &=&
 \left( {\mathcal L_2}  - \omega_2(x) \right) P_2(x,q,t)  \label{eqs:moteur+chimie} \\
&+&   \omega_1^{0}(x) P_1(x,q,t)
 +
 \omega_1^{1}(x)  P_1(x,q-1,t). \nonumber
 \end{eqnarray}
We use a notation similar to that of Ref.~\cite{pre-FT}, where
$\omega_i^l(x)$ denotes the transition rate at position $x$ from
the internal state $i$ with $l(=-1,0,1)$ ATP molecules consumed.
This leads to
\begin{eqnarray}
\omega_1^0(x) &=& \omega e^{(U_1-fx)/k_B T} \,, \\
\omega_2^0(x)  &=&  \omega e^{(U_2-fx)/k_B T}  \,,  \\
\omega_1^1(x)  &=&   \psi e^{(U_1-fx)/k_B T+\Delta \mu} \,,  \\
\omega_2^{-1}(x)   &=&   \psi e^{(U_2-fx)/k_B T}  \, .
\end{eqnarray}
 We also have
$\omega_1(x)=\omega_1^0(x)+\omega_1^1(x)$ and
$\omega_2(x)=\omega_2^0(x)+\omega_2^{-1}(x)$.
The operators ${\mathcal L_1}$  and ${\mathcal L_2}$  act on a function $\Phi$ as
 \begin{equation}
 {\mathcal L_i} =  D_0 \frac{\partial^2 \Phi  }{\partial x^2}   +
 D_0  \frac{\partial}{\partial x} \left( \frac{{U_i'}(x) - F }{k_B T}  \,\,  \Phi  \right)
 \,\,\,\, \,\,\,\,  i=1,2  \, .
 \end{equation}

As above, we introduce  two generating functions
   $F_{1,\lambda,\gamma}$ and   $F_{2,\lambda,\gamma}$,
    depending on two parameters $\lambda$ and
 $\gamma$ which are conjugate variables to the position of the ratchet
  and to the ATP counter $q$. We have for $i=1,2$,
  \begin{equation}
   F_{i, \lambda,\gamma}(\zeta, t) =
 \sum_q  \hbox{e}^{\gamma q}  \sum_n  \hbox{e}^{\lambda(\zeta +n)}
 P_i(a(\zeta +n),q; t).
\label{GFMecChim}
 \end{equation}
 The evolution equation for  these generating functions  is obtained
 from Eq.~(\ref{eqs:moteur+chimie}) as
 \begin{equation}
 \frac{\partial }{\partial t}
  \left( \begin{array}{c}
    F_{1,\lambda,\gamma} \\   F_{2,\lambda,\gamma}
   \end{array}  \right)
 =   {\mathcal  L}(\lambda,\gamma)  \left( \begin{array}{c}
    F_{1,\lambda,\gamma} \\   F_{2,\lambda,\gamma}
   \end{array}  \right)  \, ,
\label{EvolMecChim}
 \end{equation}
 with the operator  ${\mathcal  L}(\lambda,\gamma)$  decomposed as
\begin{equation}
 {\mathcal  L}(\lambda,\gamma) = {\mathcal D}(\lambda) + {\mathcal N}(\gamma),
\end{equation}
 with ${\mathcal D}(\lambda)$ the diagonal matrix
  ${\rm diag}({\mathcal L_1}(\lambda)-\omega_1,{\mathcal L_2}(\lambda)-\omega_2)$,
  where the deformed operators ${\mathcal L_1}(\lambda)$  and
 ${\mathcal L_2}(\lambda)$ have the form written  in Eq.~(\ref{def:Ldeformed})
 with  $U_{eff}(x)$ given by $ U_i(x) - Fx$ for $i=1,2$,
 respectively.
 The  operator ${\mathcal  N}(\gamma)$  is defined as
 \begin{equation}
{\mathcal  N}(\gamma)  = \left( \begin{array}{cc}
      0 &  \omega_2^0+\omega_2^{-1} e^{-\gamma}  \\
 \omega_1^0+ \omega_1^1 e^{\gamma}
 &   0
   \end{array}  \right).
\label{LdefMecChim}
\end{equation}
Consider now the diagonal matrix $Q$ defined by ${\rm diag} (
e^{-U_1/{k_B T}},e^{-U_2/{k_B T}})$.
By direct calculation, one can verify  that
$Q^{-1} {\mathcal N}(\gamma) Q = {\mathcal N^\dagger}
\left(-\Delta\mu -\gamma \right).$
From Eq.~(\ref{OperatorGC1}), one obtains $Q^{-1} {\mathcal
D}(\gamma) Q = {\mathcal D^\dagger}  \left(-\Delta\mu -\gamma
\right)$. By combining these two equations, we conclude that
\begin{equation}
 Q^{-1}  {\mathcal  L}(\lambda,\gamma) Q =
 {\mathcal  L^\dagger}
 \left(-f -\lambda, -\Delta\mu -\gamma \right),
\label{OperatorGCMecaChim}
\end{equation}
which leads to the Gallavotti-Cohen symmetry:
\begin{equation}
 \Lambda(\lambda,\gamma) =
  \Lambda\left(-f -\lambda,  -\Delta\mu -\gamma \right),
\label{FTRatchet2var}
 \end{equation}
 where $\Lambda(\lambda,\gamma)$ is the largest eigenvalue of ${\mathcal  L(\lambda,\gamma)}$.
This relation is the equivalent of Eq.~(\ref{GC for theta
meca-chimie}), which was derived for the
 discrete model. If we consider only the mechanical displacement of the ratchet,
 the relevant eigenvalue $\Theta(\lambda)$ is given by
  $\Theta(\lambda)=\Lambda(\lambda,0)$, which clearly does not satisfy the fluctuation relation
 of the form Eqs.~(\ref{GC for theta}-\ref{GC for
P})  as shown in Fig.~\ref{fig eigenvalue_switching}. In
Fig.~\ref{fig eigenvalue 2 variables}, we have computed $\Lambda(f
\eta,-\Delta \mu/2)$ for the same potentials and with rates
$\omega_i^l(x)$ of the form given above with $\omega(x)=5
\tau^{-1}$ and $\phi(x)=10 \tau^{-1}$. We have verified that in
all cases the symmetry of Eq.~(\ref{FTRatchet2var})  holds.

To conclude this section on the continuous flashing ratchet,
 we emphasize the following two points: (i) we have proved
 that the   flashing ratchet satisfies the fluctuation theorem
 without having to   adjust  any   parameter in  the system
 to enforce the validity of this theorem.  The only constraints on
 the  switching rates  were given {\it  a priori}  from thermodynamics
 and kinetic theory  and these requirements  are  always taken into account
  in  the very definition of the model (see e.g. \cite{parmeggiani}).
  What we  have shown  is that these
 thermo-kinetic  constraints are enough  to imply  the Gallavotti-Cohen symmetry,
 which itself has far reaching consequences on the model.
 (ii) In order to derive the fluctuation theorem, all relevant
 microscopic degrees of freedom must be involved. For example,
 in the flashing rachet model, the position  variable alone
 does not obey the  fluctuation theorem and the
 chemical variable that counts how many ATP molecules have been
 consumed by the motor during its  displacement has to be
 taken into account. The  Gallavotti-Cohen symmetry thus  leads
 to  global mechano-chemical constraints on the modes
 of operation of the motor.

\section{Conclusions}
A first simple and useful message to take from this study is
 that the dynamics of a molecular motor can be
described by the evolution of a random walker in an
effective potential $U_{eff}(x,y)$ where $x$ is the
mechanical variable and $y$ is the chemical variable \cite{bustamente}.
The periodicity of the potential along $x$ and $y$
implies that the potential has an egg-carton shape.

The symmetry of the fluctuation relations for the currents is valid in general for
the flashing ratchet model only when internal degrees of freedom are
taken into account. This raises a fundamental question concerning
the validity of fluctuations relations and their applicability
 to other types of ratchet models \cite{oshanin:2004,slanina:2008,reiman,hanggi}.
More generally, other  mechanisms exist which are known to produce
 deviations from fluctuations relations \cite{jarzynski_coarse_graining},
and it would be valuable to know whether fluctuations relations
can always be restored by enlarging the phase space and
by modifying the dynamics accordingly.

On the experimental side, it would be very interesting to
investigate fluctuations relations for molecular motors using
single molecule experiments, in a way similar to what was achieved
in colloidal beads or biopolymers experiments
\cite{Liphard-2002}-\cite{chetrite-2009}.
Using fluorescently labeled ATP molecules, recent experiments with
myosin 5a and with the $F_1$-ATPase rotary motor, aim at simultaneous
recording of the turnover of single fluorescent ATP molecules and
the resulting mechanical steps of the molecular motor \cite{yanagida-2007}.
These exciting results indicate that a
simultaneous measurement of the values of the mechanical
 and chemical variables  of the motor is achievable in practice, and therefore from the
 statistics of such measurements it may be possible to obtain the distribution of probability
to find the motor at a specific position and
with a specific number of molecules of ATP consumed.
With enough statistics, one could thus in principle
verify Eq.~(\ref{GC for theta meca-chimie}).
Such an experimental  verification would confirm
that the Gallavotti-Cohen symmetry is a fundamental  constraint
that plays an essential role in the mechano-chemical coupling of
molecular motors.

 Finally, besides the  Gallavotti-Cohen fluctuation theorem,
 many exact non-equilibrium relations have been discovered
 during the last decade, the most famous one  being
 the Jarzynski identity \cite{Jarzynski:1997} and its generalization by Crooks \cite{crooks}.
 These identities, originally derived for systems being driven out
of a state of thermodynamic equilibrium,
 have been  extended by Hatano-Sasa to systems prepared in
 non-equilibrium stationary states and following Markovian dynamics
  \cite{hatano-sasa:2001}.
This more general fluctuation relation leads with a proper choice of
observables to generalizations of the well-known fluctuation-dissipation
known for systems close to equilibrium
\cite{prost:2009,chetrite:2008,seifert:2006,maes:2009}. These
generalized fluctuation-response relations hold for systems prepared in
 non-equilibrium stationary states and following Markovian dynamics.
The implications of the Crooks fluctuation theorem for kinesin has been
analyzed in Ref.~\cite{calzetta-2009}, while the implications of generalized fluctuation-response
relations for molecular motors have been
considered recently \cite{lacoste-2010,seifert-2010}.

 All these various relations can be interpreted  as
 universal constraints that
 have to be obeyed by systems far from equilibrium
 regardless of their detailed structure.
 It would be of great interest to explore
  the consequences of these relations in
 the field of ratchet models (both at the single motor level and at the level
 of many motors) and to draw from them some
 measurable predictions that could be verified experimentally
 on molecular motors.

\section*{Acknowledgment}
D.L. acknowledges support from
 the Indo-French Center CEFIPRA (grant 3504-2), and
 the IIT Kanpur for hospitality during the 2010 Golden Jubilee.

\bibliographystyle{plain}

\end{document}